\documentclass[runningheads]{llncs}
\usepackage[T1]{fontenc}
\usepackage{graphicx}
\usepackage{epsfig}
\usepackage{subcaption}
\usepackage{calc}
\usepackage{amssymb}
\usepackage{amstext}
\usepackage{amsmath}
\usepackage{multicol}
\usepackage{pslatex}
\usepackage{algorithm2e}
\usepackage[bottom]{footmisc}
\usepackage{xcolor}

\usepackage{multirow}
\usepackage{tabularx}
\usepackage{bm}
\usepackage{graphicx}
\usepackage{upgreek}
\newcommand\inlineeqno{\stepcounter{equation}\ (\theequation)}
\usepackage{cancel}

\begin{document}
\title{Enforcing Soft Monotonicity Constraints for Recursive Gaussian Process Regression  in Real Time}
\titlerunning{Monotonicity for RGP}
%
\author{Ricus Husmann\thanks{0009-0006-0480-8877} \and
Sven Weishaupt\thanks{0009-0007-0601-4605} \and
Harald Aschemann\thanks{0000-0001-7789-5699}}
%
\authorrunning{Ricus Husmann et al.}
%
\institute{Chair of Mechatronics, University of Rostock, Rostock, Germany\\
\email{\{ricus.husmann, sven.weishaupt, harald.aschemann\}@uni-rostock.de}}
\maketitle              
\begin{abstract}
In this work, we introduce a real-time capable algorithm for considering monotonicity assumptions for recursive Gaussian Process regression (RGP). Therefore, we present how to efficiently calculate the RGP gradients online. Then, we utilize an extended Kalman filter and pseudo-measurements in combination with a ReLU pseudo-measurement function to enforce soft inequality constraints. This work builds upon a previously published conference paper with the same goal and a similar fundamental approach. Opposite to our previous work, however, we now use an exact covariance calculation for the RGP gradients. Furthermore, we also present  a real-time optimized  version of this algorithm with less simplifications compared to the previously published version. These and several other algorithmic innovations lead to an algorithm with greatly improved numerical robustness. The algorithm is validated and compared to its previously published version for a 2D numerical example.  The paper is concluded with a successful experimental validation of the developed algorithm for the monotonicity-preserving learning of pneumatic valve characteristics for the control of a pneumatic system, leveraging a partial input - output linearization.

\end{abstract}
\section{Introduction}
Considerable progress can be observed in the online identification of system models. A basic approach is the definition of parametric functions, such as polynomial ansatz functions, followed by the application of recursive least-squares regression as described in \cite{Blum:1957}. If models with non-measurable states or parameters have to be identified, this method can be extended using linear Kalman Filters (KF) or Unscented/Extended Kalman Filters (UKF/EKF), as proposed in \cite{Kalman:1960} and \cite{Julier:1997}. In the presence of additional inequality constraints, Moving Horizon Estimation (MHE) techniques are particularly suitable \cite{Haseltine:2005}. For more general approaches, online-capable training methods for neural networks are available, as discussed in \cite{Jain:2014}.

Gaussian Processes (GP) have been established as a popular non-parametric alternative to neural networks (NNs). They are typically more data-efficient than neural networks, robust to overfitting, and - as a main advantage compared to NNs — they provide an uncertainty quantification for the predicted values, see \cite{Schuerch:2020}. However, their non-parametric nature and the $O(n^3)$ increase in computational effort w.r.t. the number of utilized data points pose a major challenge for online implementation. Nevertheless, the literature provides suitable methods to address this issue, such as active-set methods that limit the number of utilized measurement points \cite{Quinonero:2005}. Furthermore, a promising algorithm was presented in \cite{Huber:2013} in the form of recursive Gaussian Process-regression (RGP). The main idea is to define the GPs as parametric functions-based on user - defined basis vectors, thereby preserving many benefits of GP regression while maintaining a low computational load.

For many modelling tasks, a certain amount of prior knowledge is available. This may include bounds on model outputs or monotonicity assumptions. Such knowledge may originate from physical properties or appear in the form of stability-preserving constraints in a control setting. Incorporating this information during learning has the potential to yield superior models with significantly less data. In neural networks, such assumptions can be considered by modifying the loss function, as in \cite{Cuomo:2022}. For standard GPs, several methods exist to incorporate prior knowledge, for example regarding system structure \cite{Beckers:2022} or inequality constraints \cite{Veiga:2020}. To the best of our knowledge, however, the integration of inequality constraints into recursive Gaussian Processes represents a novel development.

\vspace{0.2 cm}
\noindent The main contributions of the paper are:
\begin{itemize}
\item Computationally efficient consideration of monotonicity constraints w.r.t. recursive Gaussian Processes in an EKF update
\item Real-time optimized  version of this algorithm
\item Real-time implementation and utilization of the presented algorithm within the control structure for a pneumatic valve.
\end{itemize}
\vspace{0.2 cm}

In \cite{RGP_mon:2025}, we presented a similar algorithm with the same goal of considering monotonicity constraints for RGPs. There we utilized a simplified covariance prediction. While the real-time capable version of that algorithm tends to work very well in practice, there are certain cases, where it may lead to numerical instability. This is one of the points which we address in this work by a new exact covariance prediction of the gradients. The new contributions compared with \cite{RGP_mon:2025} are thus:
\begin{itemize}
\item Monotonicity constraints under usage of exact covariance predictions for the RGP-gradients with an efficient Cholesky decomposition-based implementation
\item New real-time-optimized  version  with less simplifications
\item Numerical validation for a 2D example and comparison with the previously published version
\item New and deeper experimental validation on a pneumatic test rig
\end{itemize}

The paper is structured as follows: First, we recapitulate the RGP in Sec.~\ref{sec:RGP}. Then, the calculation of the RGP gradients is presented in Sec.~\ref{sec:gradient}. In Sec.~\ref{sec:EKF} we derive our general approach to handle inequality constraints with an EKF structure and show how this can be applied to monotonicity constraints. In Sec.~\ref{sec:realTime}, we then present a version of this algorithm which is optimized for a real-time implementation. After a statistical evaluation and comparison in Sec.~\ref{sec:simul_eval}, we then experimentally validate the derived real-time optimized algorithm and its application within a model-based controller on a pneumatic test rig in Sec. \ref{sec:ex_val}.  The paper finishes with a conclusion and an outlook.

\section{Recursive Gaussian Process Regression}
\label{sec:RGP}
In this chapter, we briefly describe our implementation of the recursive Gaussian Process regression (RGP) from \cite{Huber:2013}. 

At timestep $k$, we assume a scalar measurement $y_k=y(k)$ of a constant hidden function of the following form
\begin{equation}
y_k=z(\bm{\zeta}_k)+ \epsilon_k
\end{equation}
with the Gaussian white measurement noise  $\epsilon_k \sim  \mathcal{N} (0,\sigma_y)$  and  the deterministic inputs $\bm{\zeta}_k \in \mathbb{R}^{n_z}$ to the function.  The RGP algorithm is used to learn a finite dimensional RGP model $\tilde{z}_k(\bm{\zeta}_k)$ for the hidden function $z$ by utilizing the noisy measurements $y_k$. The RGP model can be used to provide the mean value  $\mu_{k}^p=\mathrm{E}\{\tilde{z}_k(\bm{\zeta}_k)\}$ and variance predictions ${c}_{k}^p=\mathrm{Var}\{\tilde{z}_k(\bm{\zeta}_k)\}$ for given $\bm{\zeta}_k$.

As usual, we utilize a Squared Exponential (SE) kernel, which for some matrices $\bm{X}$ and $\bm{X}'$ is defined as
\begin{equation}
k(\bm{X},\bm{X}')=\sigma_K^2 \cdot \exp(-(\bm{X}-\bm{X}')^T(\bm{X}-\bm{X}')(2 L)^{-1}) \,,
\end{equation}
and a zero mean function. For a particular set of input matrices 
\begin{equation}
\bm{X}=\begin{bmatrix}
\chi_{1,1} & \chi_{1,2}\\
\chi_{2,1} & \chi_{2,2}\\
\chi_{3,1} & \chi_{3,2}\\
\end{bmatrix} \,,
\bm{X}'=\begin{bmatrix}
\chi_{1,1}' & \chi_{1,2}'\\
\chi_{2,1}'& \chi_{2,2}'\\
\end{bmatrix}
\end{equation}
the respective kernel matrix is, e.g., defined as follows
\begin{equation}
\bm{K}=k(\bm{X},\bm{X}')= \begin{bmatrix}
k([\chi_{1,1},\chi_{1,2}],[\chi_{1,1}',\chi_{1,2}']) & k([\chi_{1,1},\chi_{1,2}],[\chi_{2,1}',\chi_{2,2}']) \\
k([\chi_{2,1},\chi_{2,2}],[\chi_{1,1}',\chi_{1,2}']) & k([\chi_{2,1},\chi_{2,2}],[\chi_{2,1}',\chi_{2,2}']) \\
k([\chi_{3,1},\chi_{3,2}],[\chi_{1,1}',\chi_{1,2}']) & k([\chi_{3,1},\chi_{3,2}],[\chi_{2,1}',\chi_{2,2}']) \\
\end{bmatrix} \,.
\end{equation}

In our application, the hyperparameters $L$ and $\sigma_K$ are user-defined. As elaborated in Subsec.~\ref{sec:Inp_Norm}, a joint length scale $L$ is defined for all input dimensions, and the possibly different input ranges are addressed by an extra normalization step with the normalization function $\bm{f}_{norm}(\bm{\zeta}_k)$, as derived in  Subsec.~\ref{sec:Inp_Norm}.  As also detailed in Subsec.~\ref{sec:Inp_Norm}, $\bm{X}\in \mathbb{R}^{N_{X}\times n_z}$ refers to the user defined constant basis vectors, which are defined during initialization, and $\bm{\chi}_k\in \mathbb{R}^{n_z}$ denotes the current test input, which are the normalized hidden function inputs $\bm{\zeta}_k$. The mean values $\bm{\mu}^g_{k}$ and the covariance matrix $\bm{C}_{k}^g$ of the kernels, which are updated recursively, give the RGP algorithm a KF-like structure.

The following variables can be precalculated offline:
 \begin{center}
\begin{tabularx}{1.028\columnwidth}{l X r}  
  \multirow{3}{1em}{\rotatebox{90}{{Offline}}} 
  & $\bm{K}=k(\bm{X},\bm{X})$,  &$\inlineeqno $ \\ 
& $\bm{\mu}_{0}^g= \bm{0}$ ,&\\ 
& $\bm{C}_{0}^g= \bm{K} \,.$ &\\  
\end{tabularx}
\end{center}
Given a zero mean function and a single measurement per timestep, the prediction or inference step simplifies to:
\begin{center}
\begin{tabularx}{1.028\columnwidth}{l X r} 
  \multirow{4}{1em}{\rotatebox{90}{{Inference}}} & $\bm{\chi}_k=\bm{f}_{norm}(\bm{\zeta}_k) $  ,&$\inlineeqno$\\
  & $\bm{j}_{k}^T =k(\bm{\chi}_k^T,\bm{X}) / \bm{K} $  ,&\\ 
& $\mu_{k}^p=\bm{j}_k^T  \bm{\mu}^g_{k} $,& \\ 
& $c_{k}^p=\underbrace{k(\bm{X}_k,\bm{X}_k)}_{\sigma_K^2}+\bm{j}_k^T (\bm{C}_{k}^g-\bm{K}) \bm{j}_k \,,$ & \\  
\end{tabularx}
\end{center}
where the superscript $p$ indicates the RGP prediction for the test inputs $\bm{\chi}_k$. This prediction is used in the following update step:
\begin{center}
\begin{tabularx}{1.028\columnwidth}{l X r} 
  \multirow{3}{1em}{\rotatebox{90}{{Update}}} & $\bm{g}_{k}=\bm{C}_{k}^g \bm{j}_k (c_{k}^p+ \sigma_{y}^2 )^{-1}$ , &$\inlineeqno$ \\ 
& $\bm{\mu}_{k+1}^g= \bm{\mu}_{k}^g+\bm{g}_{k}  (y_{k}- \mu_{k}^p) $,& \\ 
& $\bm{C}_{k+1}^g=\bm{C}_{k}^g-\bm{g}_{k} \bm{j}_k^T\bm{C}_{k}^g \,.$ & \\ 
\end{tabularx}
\end{center}

The $/$ in the inference step   refers to a solution of the linear matrix equation. In the original work of \cite{Huber:2013}, the algorithm uses an offline-precomputed inverse of the Kernel matrix $\bm{K}$. While this method is computationally very efficient, it did not prove to be numerically stable since $\bm{K}$ is  ill-conditioned in many cases. In \cite{RGP_dKF:2025}, we thus proposed an online solution-based on an offline QR decomposition, which is also used in this work.

\subsection{Input Normalization} 
\label{sec:Inp_Norm}
We define the basis vectors for all input axes as an equidistant grid with step size $1$. This leads to a matrix of size $ N_{X}\, \times \, n_{z}$, with $N_{X}=\left(\prod_{i=1}^{n_{z}} N_i \right)$, that contains all vertices of the grid, where $n_{z}$ denotes the input dimension of the RGP, and $N_i$ the number of points in the respective dimension. As an example  $n_z=2$,  $N_1=2$ and $N_2=3$ lead to the basis vectors
\begin{equation}
\bm{X}=\left[\begin{matrix}  \bm{\chi}_1^T\\ \bm{\chi}_2^T\end{matrix}\right]^T=\left[\begin{matrix}  0 &1 &0&1&0&1 \\  0 &0 &1&1&2&2 \end{matrix}\right]^T \,.
\end{equation}
To address the ranges of the actual inputs $\zeta_{i,k}$ in each dimension, we introduce the normalization step
\begin{equation}
\chi_{i,k}=f_{norm}(\zeta_{k,i})=(\zeta_{i,k}-\underline{\zeta}_{i}) \cdot \underbrace{\frac{N_i-1}{\overline{\zeta}_{i}-\underline{\zeta}_{i}}}_{\beta_i}  \,, \label{eq:normalization}
\end{equation}
which is applied before each RGP evaluation. Here, $\underline{\zeta}_{i}$ and $\overline{\zeta}_{i}$ denote the corresponding lower and upper bounds of the input, and $\beta_i$ is a constant factor, which is used in Subsec.~\ref{sec:gradient}. 

The normalization and the use of a joint length $L$ was originally introduced to handle numerical issues that may occur for large $L$ in the standard inversion-based RGP formulation. With the normalization step, a universal maximum $L_{max}$ -- independent of the system -- could be determined to maintain numerical stability. Whereas the issue of numerical instability for large $L$ no longer applies, the normalization reduces the number of free hyperparameters  and makes it easier to find good hyperparameters for new systems. Consequently, it is also used in this work.

\section{RGP Gradients}
\label{sec:gradient}
In this section, we present the calculation of the mean - values 
\begin{equation}
\bm{\mu}_{m,k,i}=\mathrm{E}\left\{\left.\frac{\partial \tilde{z}_k}{\partial {\zeta}_{i,k}}\right|_{\tilde{\bm{X}}}\right\}
\end{equation} 
and covariances 
\begin{equation}
\bm{C}_{m,k,i}=\mathrm{Cov}\left\{\left.\frac{\partial \tilde{z}_k}{\partial {\zeta}_{i,k}} \right|_{\tilde{\bm{X}}}\right\}
\end{equation} 
of the RGP gradients regarding dimension $i$ evaluated on a test grid $\tilde{\bm{X}} \in \mathbb{R}^{\tilde{N}_{X}\, \times \, n_{z}}$ with $\tilde{N}_{X}=\left(\prod_{i=1}^{n_{z}} \tilde{N}_i \right)$ and the test grid size $\tilde{N}_i$ for the respective dimension.  The gradients will later be used to enforce monotonicity constraints. 

\begin{figure}
	 \begin{center}
		 \includegraphics[width=1\linewidth]{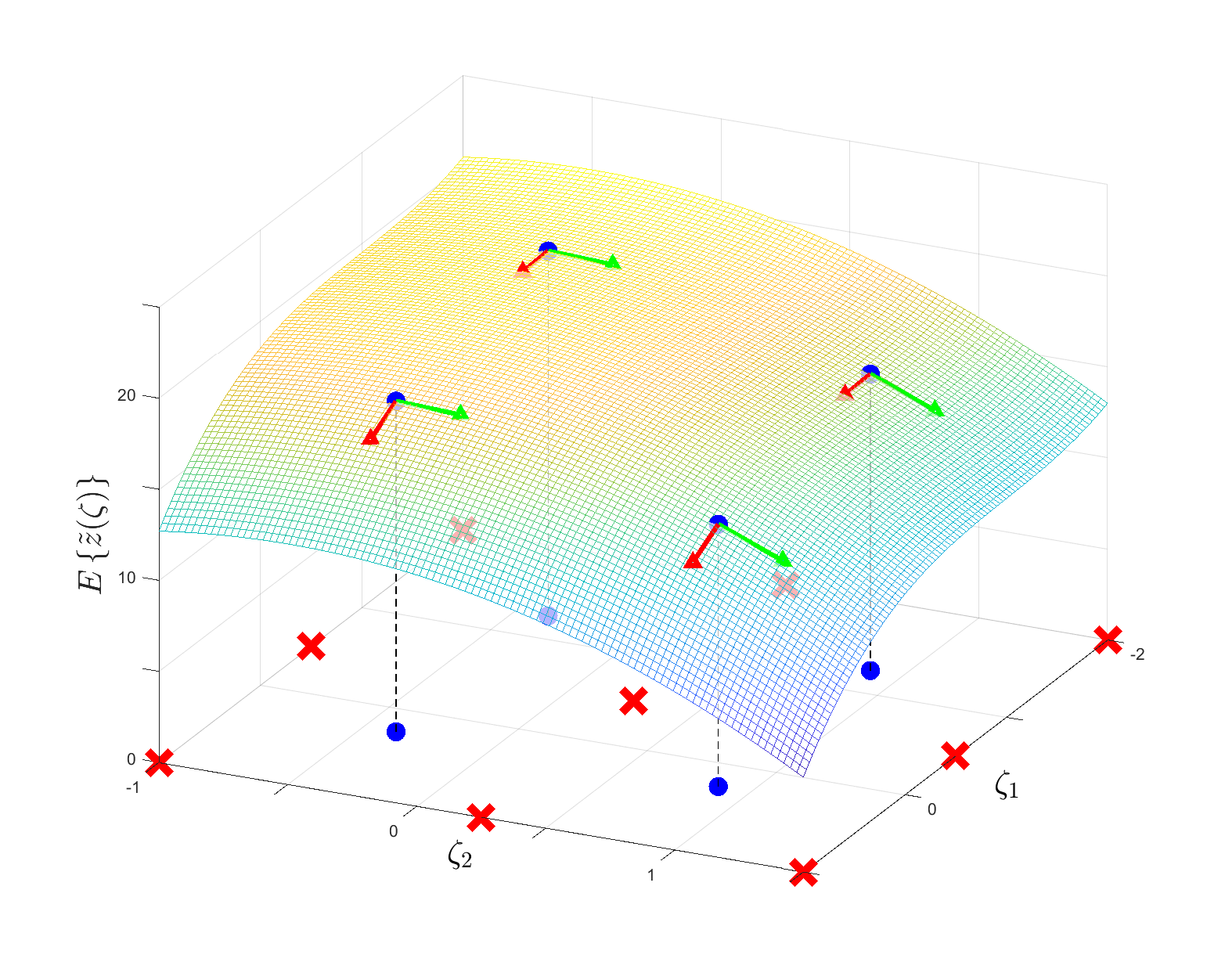}
		  \caption{Exemplary gradients for a 2D RGP. The basis vector grid $\bm{X}$  is denoted by red crosses. The test grid $\bm{\tilde{X}}$ is given by the blue dots (both depicted in $\zeta$-coordinates). The red arrows indicate the gradients $\mu_{m,k,1}$ for the test grid in direction of dimension $i=1$ and the green arrow the gradients $\mu_{m,k,2}$ for dimension $i=2$ respectively.} 
		  \label{pic:TwoDGridExample}
	 \end{center}
\end{figure}

In Fig. \ref{pic:TwoDGridExample}, we provide a sketch of a 2D RGP with a basis vector grid of size ${N}_1={N}_2=3$ as well as a gradient test grid of size $\tilde{N}_1=\tilde{N}_2=2$. Both are depicted in $\zeta$-coordinates. Furthermore, the mean values of the gradients $\bm{\mu}_{m,k,i}$ in direction of the respective input dimension $i$  are depicted for the test grid.

\subsection{Single Directional Gradient}
As derived, for example, in \cite[pp.185-192]{mchutchon2:2015}, the mean-value gradient of the RGP prediction with SE kernels w.r.t.~to the $i$-th input dimension is provided by the function
\begin{equation}
\bm{\mu}_{m,i}=\underbrace{\left(-\frac{\beta_i}{L} \left(\left(\bm{\tilde{\chi}}_{i}-\bm{\chi}_i^T\right)\odot k(\bm{\tilde{X}},\bm{X})\right)/\bm{K} \right)}_{\bm{H}_{m,i}} \bm{\mu}^g_{k}, \label{eq:gradient}
\end{equation}
where $\odot$ denotes the Schur- or Hadamard product, and $\tilde{\bm{\chi}}_i$ and $\bm{\chi}_i$ are the respective columns of the test input grid and basis vector grid adequate to the current input dimension  $i$. The constants $\beta_i$ arise from the normalization as becomes clear in \eqref{eq:normalization}. 

Also following  \cite[pp.185-192]{mchutchon2:2015} the covariance of the RGP prediction can be calculated accordingly by evaluating 
\begin{equation}
\bm{C}_{m,i,k}=\bm{H}_{m,i} \left(\bm{C}_{k}^g-\bm{K} \right)\bm{H}_{m,i} ^T+ \bm{R}_{m,i} 
\end{equation}
with 
\begin{equation}
\bm{R}_{m,i}= \frac{1-\left((\bm{\tilde{\chi}}_{i}-\bm{\tilde{\chi}}_{i}^T)\odot (\bm{\tilde{\chi}}_{i}-\bm{\tilde{\chi}}_{i}^T)\right)/L^2}{L^2} \odot k(\bm{\tilde{X}},\bm{\tilde{X}}) \cdot \beta_i^2 \,. \label{eq:grad_diag}
\end{equation}

\subsection{Multi - Directional Gradient}
\label{sec:simul_eval}
In general, the gradients of an GP or RGP w.r.t. multiple dimensions evaluated for a grid are correlated. If we combine the RGP gradient prediction for all dimensions into one operation, the mean values can be written  as follows 

\begin{equation}
\bm{\mu}_{m,k}=\bm{H}_{m} \bm{\mu}^g_{k}, \label{eq:gradientd1}
\end{equation}
with 
\begin{equation}
\bm{\mu}_{m,k}=\mathrm{E}\left\{\left[\left.\frac{\partial \tilde{z}_k}{\partial {\zeta_{1,k}}}\right|_{\bm{\tilde{X}}}^T,\left.\frac{\partial \tilde{z}_k}{\partial {\zeta_{2,k}}}\right|_{\bm{\tilde{X}}}^T,\dots\right]^T\right\}
\end{equation}
and 
\begin{equation}
 \bm{H}_{m}=\left[ \bm{H}_{m,1}^T,\bm{H}_{m,2}^T, \cdots \bm{H}_{m,n_z}^T\right]^T \,, \label{eq:meas_matr}
\end{equation}
where $\bm{H}_{m,i}$ are given by \eqref{eq:gradient}. As there are now $\tilde{N}_X\cdot n_z$  predicted gradients, the "measurement"   matrix is of dimension $\bm{H}_{m} \in \mathbb{R}^{(\tilde{N}_X\cdot n_z) \times \tilde{N}_X}$.

The complete covariance prediction of the RGP gradients might be thus written as
\begin{equation}
\bm{C}_{m,k}=\bm{H}_{m} \left(\bm{C}_{k}^g-\bm{K} \right)\bm{H}_{m} ^T+  \bm{R}_{m} \,,
\end{equation}
with 
\begin{equation}
\bm{C}_{m,k}=\mathrm{Cov}\left\{\left[\left.\frac{\partial \tilde{z}_k}{\partial {\zeta_{1,k}}}\right|_{\bm{\tilde{X}}}^T,\left.\frac{\partial \tilde{z}_k}{\partial {\zeta_{2,k}}}\right|_{\bm{\tilde{X}}}^T,\dots\right]^T\right\}
\end{equation}
and
\begin{equation}
\bm{R}_{m}=
\begin{bmatrix}
\bm{R}_{m,1} & \bm{R}_{m,12} & \bm{R}_{m,13} & \cdots & \bm{R}_{m,1 n_z} \\
\bm{R}_{m,12} & \bm{R}_{m,2} & \bm{R}_{m,23} & \cdots & \bm{R}_{m,2 n_z} \\
\bm{R}_{m,13} & \bm{R}_{m,23} & \bm{R}_{m,3} & \cdots & \bm{R}_{m,3 n_z} \\
\vdots & \vdots & \vdots & \ddots & \vdots \\
\bm{R}_{m,1 n_z} & \bm{R}_{m,2 n_z} & \bm{R}_{m,3 n_z} & \cdots & \bm{R}_{m,n_z}
\end{bmatrix} \,. \label{eq:meas_cov}
\end{equation}
The diagonal elements are calculated by \eqref{eq:grad_diag}, and the off-diagonal elements are  given by
\begin{equation}
\bm{R}_{m,i_1 i_2}= \frac{(\bm{\tilde{\chi}}_{i_1}-\bm{\tilde{\chi}}_{i_1}^T)\odot (\bm{\tilde{\chi}}_{i_2}-\bm{\tilde{\chi}}_{i_2}^T)}{L^4} \odot k(\bm{\tilde{X}},\bm{\tilde{X}}) \cdot \beta_{i_1}\cdot \beta_{i_2} \,. \label{eq:CrossCov}
\end{equation}

\subsection{Real - Time Implementation}
\label{sec:SpeedUp1}
The test vector grid $\bm{\tilde{X}}$ is assumed to be constant during runtime. In the recursive Gaussian process regression, hence, only $\bm{\mu}^g_{k}$ and $\bm{C}_{k}^g$ change during runtime, whereas $\bm{H}_{m}$ and $\bm{R}_{m}$ can be precomputed offline. Consequently, the computation of the mean value of the gradient for the test-vector grid might be written as
\begin{equation}
\bm{\mu}_{m,k}=\underbrace{\bm{H}_{m}}_{const.} \bm{\mu}_k^g \,.
\end{equation}
Since $\bm{K}$ is also precomputed for the RGP, the covariance calculation of the gradient is equally mostly dependent on constant matrices
\begin{equation}
\bm{C}_{m,k}=\underbrace{\bm{H}_{m}}_{const.} \bm{C}_{k}^g\underbrace{\bm{H}_{m} ^T}_{const.}+\underbrace{\bm{R}_{m}- \bm{H}_{m} \bm{K} \bm{H}_{m} ^T}_{\bm{R}_{m,ges}=const.} \,.\label{eq:Cgrad_ges}
\end{equation}

\section{Enforcing Monotonicity Constraints for RGPs}
\label{sec:EKF}
In this chapter, we present our implementation to enforce (soft) monotonicity constraints for RGPs. The proposed method is based upon an EKF update for inequality constraints, which is described in the sequel after the precise problem formulation. Afterwards, we present the formulation of RGP monotonicity as a constraint. We put emphasis on a computational speedup of the algorithm in the next subsection and summarize the complete algorithm. A further speedup is provided by a real-time optimized version of the algorithm in the following Sec. \ref{sec:realTime}.

We assume previous knowledge of the monotonicity of the hidden function $z(\bm{\zeta}_k)$ w.r.t. its inputs $\bm{\zeta}_k$, which can be stated in an inequality constraint regarding the partial derivatives $\frac{\partial z}{\partial \zeta_{i,k}}\gtrless 0$. To enable safety margins, this is generalized to $\frac{\partial z}{\partial \zeta_ {i,k}}\gtrless B_i$, where $B_i$ is a constant characterizing the boundary of the constraint. For simplification of the algorithm description, we always consider that all dimensions are subject to monotonicity assumptions. If no gradient information is available, one could set $B_i$ to very large values so that the respective monotonicity assumption never becomes active.

\subsection{EKF Update for Inequality Constraints}
\label{sec:Ineq_Constr}
The direct consideration of hard inequality constraints (IC) on Gaussian variables leads to truncated Gaussians, see \cite{Tully:2011}. For univariate Gaussians, the resulting mean and covariance can be calculated efficiently. For multivariate Gaussians and inequality constraints that dependent on several Gaussian input variables, however, exact solutions usually necessitate numerical methods. Here, \cite{Simon:2006} provides an overview and also discusses the use of equality constraints as exact pseudo-measurements within a KF update. This is related, however, to some numerical issues since exact measurements lead to rank-deficient updates in a KF. Alternative soft constraints, where pseudo-measurements are considered with a small uncertainty, are not subject to this problem. In this paper, hence, we take advantage of this approach and extend it towards inequality constraints.

\begin{figure}
	 \begin{center}
		 \includegraphics[width=0.6\linewidth]{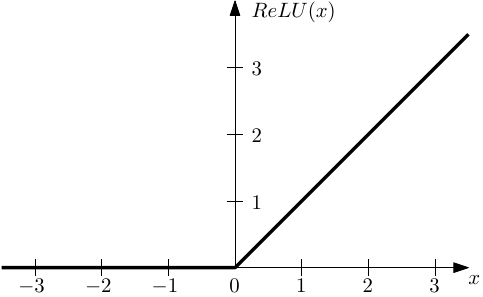}
		  \caption{Rectified linear unit (ReLU) function.} 
		  \label{pic:ReLU}
	 \end{center}
\end{figure}

The algorithm can be viewed as an extension of the previously mentioned method to enforce soft equality constraints with KF as pseudo-measurements. The extensions consist of the usage of an ReLU measurement function as depicted in Fig. \ref{pic:ReLU} and the subsequent EKF update. In the case of an inactive IC in the current step, the ReLU function in combination with the EKF "hides" the IC in the update. If, on the other hand, the IC is active in the current step, the ReLU function has no effect and the IC is considered as an equality constraint. Here, some parallels to the active-set method for constrained optimization can be drawn, see \cite{Nocedal:2006}.  Similar parallels are also drawn in \cite{Gupta:2007}, however, in combination with projection and gain-limiting methods instead of pseudo-measurements. Of course, a truncated Gaussian may differ quite dramatically in shape from a Gaussian distribution. As a result, this linearization-based approach may cause large covariance errors . To rule out corresponding covariance under-approximation, the overall covariance update related to the EKF inequality constraint is discarded at the end, as described later. This measure contributes to the "softness" of the constraints.

To simplify the implementation, we standardize all inequalities $j$ by means of the sign indicator variable $s_j$: $\hat{y}_{IC,1}<B_1 \equiv s_1\cdot (\hat{y}_{IC,1}-B_1)<0$ with $s_1=1$, $\hat{y}_{IC,2}>B_2\equiv s_2 \cdot (\hat{y}_{IC,2}-B_2)<0$ with $s_2=-1$. This corresponds to linear inequalities of the type $s_j (\bm{h}_{IC,j}^T \bm{x}_k-B_j)<0$, where $\bm{x}_k$ denotes the state vector. Now, we introduce the nonlinear measurement function, which is evaluated with the mean values
\begin{equation}
\hat{y}_{IC,j}=\tilde{h}_{IC,j}(\bm{x}_k=\bm{\mu}_{k+1}^g)=\mathrm{ReLU} (s_j (\bm{h}_{IC,j}^T\bm{\mu}_{k+1}^g-B_j))  \,.
\end{equation}
 Due to the standardization of the inequalities, all the pseudo-measurements become $y_{IC,j}=0$. The measurement functions can be concatenated in the following vector $\bm{\tilde{h}}_{IC}=\left[{\tilde{h}}_{IC,1},{\tilde{h}}_{IC,2},..\right]^T$. 

The partial derivative of the measurement, which is needed for the EKF update, is given by
\begin{equation}
\bm{\hat{h}}_{IC,j,k}^T = \left(\left.\frac{\partial \tilde{h}_{IC,j}(\bm{x}_k)}{\partial \bm{x}_k}\right|_{\bm{x}_k=\bm{\mu}_{k+1}^g}  \right)^T  
=\left\{ 
  \begin{array}{ c l }
    s_j  \bm{h}_{IC,j}^T & \quad \mathrm{if}\quad s_j (\bm{h}_{IC,j}^T\bm{\mu}_{k+1}^g-B_j)>0 \\
    \left[0,0,..\right]& \quad \mathrm{otherwise}, \\
  \end{array}
\right. \,,
\end{equation}
which can be concatenated as well to the following linearized measurement matrix $\bm{\hat{H}}_{IC,k} = \left[\bm{\hat{h}}_{IC,1,k}^T,\bm{\hat{h}}_{IC,2,k}^T,..\right]^T$. The update can then be computed as in a standard EKF, with the pseudo-measurements $y_{IC,j}=0$, according to
\begin{align}
{\bm{\tilde{G}}}_{k}&=\bm{C}_{k+1}^g \bm{\hat{H}}_{IC,k}^T ({\bm{\hat{H}}_{IC,k}}  \bm{C}_{k+1}^g {\bm{\hat{H}}_{IC,k}}^T+ \bm{R}_{IC})^{-1} ~, \\  \nonumber
 \bm{\mu}_{k+1}^c&= \bm{\mu}_{k+1}^g-{\bm{\tilde{G}}}_{k} \cdot \bm{\tilde{h}}_{IC}(\bm{\mu}_{k+1}^g) ~, \\ \nonumber
 \bm{C}_{k+1}^c&=\bm{C}_{k+1}^g-{\bm{\tilde{G}}}_{k} \bm{\hat{H}}_{IC,k} \bm{C}_{k+1}^g, ~ 
\end{align}
where $\bm{R}_{IC}$ is the pseudo-measurement noise matrix. The superscript $c$ denotes the constrained mean values and covariance.

\subsection{RGP Gradients as Inequality Constraints}
\label{sec:Grad_IC}
As discussed in Subsec. \ref{sec:SpeedUp1}, the exact prediction of the mean value of the RGP gradient for a constant grid $\tilde{\bm{X}}$ might be represented in the following form
\begin{equation}
\bm{\mu}_{m,k}=\bm{H}_{m}\bm{\mu}_k^g \,,
\end{equation}
with a constant matrix $\bm{H}_{m}$, whereas the covariance matrix prediction can be written as
\begin{equation}
\bm{C}_{m,k}=\bm{H}_{m} \bm{C}_{k}^g\bm{H}_{m} ^T +\bm{R}_{m,ges} \,,
\end{equation}
with an equally constant matrix  $\bm{R}_{m,ges}$ .

Obviously, the gradient prediction for a constant grid is linear w.r.t. the Gaussian variables of the RGP, i.e., mean values $\bm{\mu}_k^g$ and covariance matrix $\bm{C}_{k}^g$. Within the EKF pseudo-measurement update, the matrix $\bm{R}_{m,ges}$  structurally corresponds to measurement noise and complies with the structure used in Subsec. \ref{sec:Ineq_Constr}. Thus, the integration of IC regarding the RGP gradients is  straightforward with $\bm{R}_{IC}=\bm{R}_{m,ges} +\tilde{\bm{R}}_{IC}$  as well as $\bm{h}_{IC,j}^T$, which now refers to the row $j$ in $\bm{H}_m$. As in the sequel only one boundary and sign indicator variable applies per input dimension, we will refer to them in the following as $B_i$ and $s_i$. The matrix $\tilde{\bm{R}}_{IC}$ can be used to parameterize the uncertainty of this update.

\subsection{Speedup}
There are several  measures that can speedup the online evaluation of the gradient update. Since the covariance-update of the constraints is discarded, as discussed earlier, it does not need to be evaluated in the first place.

Firstly, we compute the mean values of the gradients for each direction only once $\bm{\mu}_{m,i,k}=\bm{H}_{m,i} \bm{\mu}_{k+1}^g$. Then, we introduce the gradient difference for each dimension $\bm{\Delta y_{i,k}} =\bm{\mu}_{m,i,k}-B_i $.
We can now efficiently encode the activation and deactivation action of the ReLU function by computing a diagonal activation matrix for each dimension $\bm{S}_{i,k}=diag( s_i \bm{\Delta y_{i,k}}>0) $ and calculating the currently active measurement matrix $\hat{\bm{H}}_{IC,k} = blkdiag([\bm{S}_{1,k},\bm{S}_{2,k},\dots|) \bm{H}_{m}$ as well as the respective  pseudo-measurement vector $\bm{\Delta \hat{y}_k} = \newline blkdiag( \left[\bm{S}_{1,k},\bm{S}_{2,k},\dots \right]) \left[\bm{\Delta y_{1,k}}^T, \bm{\Delta y_{2,k}}^T \dots\right]^T$. The diagonal elements of $\bm{S}_{i,k}$ are either one, if the monotonicity constraints of the respective grid point are violated by the mean value for timestep $k$, or zero otherwise.  

The mean - value update can now be written as follows
\begin{align}
\tilde{\bm{C}}_{m,k}^p&= \bm{\hat{H}}_{IC,k} \bm{C}_{k+1}^g\bm{\hat{H}}_{IC,k} ^T+\bm{R}_{m,ges}+ \tilde{\bm{R}}_{IC} \,,\\
\bm{\mu}_{k+1}^c&=\bm{\mu}_{k+1}^g-\bm{C}_{k+1}^g\bm{\hat{H}}_{IC,k}^T (\tilde{\bm{C}}_{m,k}^p)^{-1} \bm{\Delta \hat{y}_k} \,.
\end{align}
Depending on the test grid size, the positive definite matrix $\tilde{\bm{C}}_{m,k+1}^p \in \mathbb{R}^{(\tilde{N} \cdot n_z)\times((\tilde{N} \cdot n_z))}$ may become quite large. Thus, a  more efficiently and numerically robust implementation leveraging a Cholesky decomposition as well as a  solution of the resulting linear equations is beneficial
\begin{align}
\tilde{\bm{C}}_{m,k}^p&= \bm{\hat{H}}_{IC,k} \bm{C}_{k+1}^g\bm{\hat{H}}_{IC,k} ^T+\bm{R}_{m,ges}+ \tilde{\bm{R}}_{IC}\,,\\
\bm{L}_k&=chol(\tilde{\bm{C}}_{m,k}^p)\,,\\
\bm{M}_k&=\bm{L}_k/\bm{\Delta \hat{y}_k} \,,\\
\bm{\mu}_{k+1}^c&=\bm{\mu}_{k+1}^g-\bm{C}_{k+1}^g\bm{\hat{H}}_{IC,k}^T (\bm{L}^T/ \bm{M}_k) \,.
\end{align}
Here, the linear equations are solved exploiting the lower triangular structure of  $\bm{L}_k$.

\subsection{Summary: Complete Monotonicity Update}
In the following, we depict the complete algorithm for  RGP subject to a pseudo-measurement update step for the monotonicity constraints. 

Define: 
\begin{itemize}
\item Basis vectors $\bm{X}$ with the expected input bounds $\underline{\zeta}_{i}$ and $\overline{\zeta}_{i}$, the grid points (grid resolution  $N_i$  per input dimension $i$  as described in \eqref{eq:normalization}) and the respective normalization gain $\beta_i$.
\item RGP hyperparameters: length-scale $L$, vertical hyperparameter $\sigma_K$, and the measurement noise $\sigma_y$.
\item Monotonicity test vectors $\bm{\tilde{X}}$ with grid resolution  $\tilde{N}_i$  per input dimension $i$  as described in Sec. \ref{sec:gradient} .
\item Pseudo measurement noise $\tilde{\bm{R}}_{IC}$.
\item Desired boundaries $B_i$ and sign - indicator variables for the respective constraints, e.g. $s_i=-1$ for $\frac{\partial z}{\partial \zeta_{i,k}}>B_i$ and $s_i=1$ for $\frac{\partial z}{\partial \zeta_{i,k}}<B_i$.
\end{itemize}

Initialize:
\begin{center}
\begin{tabularx}{1.028\columnwidth}{l X r} 
  \multirow{6}{1em}{\rotatebox{90}{{Offline}}} & $\bm{K}=k(\bm{X},\bm{X}) $ ,  &$\inlineeqno$\\ 
& $\bm{\mu}_{0}^c=\bm{0}^T$ ,&\\ 
& $\bm{C}_{0}^g=\bm{K} $ .&\\ 
& $\bm{H}_{m}$  as given by  \eqref{eq:gradient} and \eqref{eq:meas_matr} .&\\ 
& $\bm{R}_{m,ges}$  as given by \eqref{eq:meas_cov}, \eqref{eq:CrossCov} and \eqref{eq:Cgrad_ges} .&\\ 
& $\bm{R}_{IC}=\bm{R}_{m,ges} +\tilde{\bm{R}}_{IC}$  .&\\ 
\end{tabularx}
\end{center}
Now, evaluate the following recursive algorithm for all steps $k=0,1,2,..$: \newline
1. Complete the RGP Inference  
\begin{center}
\begin{tabularx}{1.028\columnwidth}{l X r} 
  \multirow{4}{1em}{\rotatebox{90}{{Inference}}} & $\bm{\chi}_k=\bm{f}_{norm}(\bm{\zeta}_k) $  ,&$\inlineeqno$\\ 
& $\bm{j}_k^T=k(\bm{\chi}_k^T,\bm{X}) / \bm{K}$ ,& \\ 
& $ \mu_{k}^p=\bm{j}_k^T  \bm{\mu}^c_{k}$ ,&\\ 
& $c_{k}^p=\sigma_K^2+\bm{j}_k^T (\bm{C}_{k}^g-\bm{K}) \bm{j}_k $ .& \\  
\end{tabularx}
\end{center}
2. Calculate the RGP Update
\begin{center}
\begin{tabularx}{1.028\columnwidth}{l X r} 
  \multirow{3}{1em}{\rotatebox{90}{{Update}}} & $\bm{g}_{k}=\bm{C}_{k}^g \bm{j}_k \cdot(c_{k}^p+ \sigma_{y}^2 )^{-1}$ , &$\inlineeqno$ \\ 
& $\bm{\mu}_{k+1}^g= \bm{\mu}_{k}^c+\bm{g}_{k}  (y_{k}- \mu_{k}^p) $,& \\ 
& $\bm{C}_{k+1}^g=\bm{C}_{k}^g-\bm{g}_{k} \bm{j}_k^T\bm{C}_{k}^g \,.$ & \\ 
\end{tabularx}
\end{center}
3. Perform the Monotonicity Update
\begin{center}
\begin{tabularx}{1.028\columnwidth}{l X r} 
  \multirow{9}{1em}{\rotatebox{90}{{Monotonicity Update}}}&   $\bm{\mu}_{m,i,k}=\bm{H}_{m,i} \bm{\mu}_{k+1}^g$ ,&$\inlineeqno $ \\
& $\bm{\Delta y_{i,k}} =\bm{\mu}_{m,i,k}-B_i $ ,& \\ 
& $\bm{S}_{i,k}=diag( {s}_{i}  \bm{\Delta y_{i,k}}>{0})$ ,& \\ 
& $\hat{\bm{H}}_{IC,k}=blkdiag([\bm{S}_{1,k},\bm{S}_{2,k},\dots|) \bm{H}_{m}$ ,& \\ 
& $\bm{\Delta \hat{y}_k} =  blkdiag( \left[\bm{S}_{1,k},\bm{S}_{2,k},\dots \right]) \left[\bm{\Delta y_{1,k}}^T, \bm{\Delta y_{2,k}}^T \dots\right]^T$ ,& \\
& $\tilde{\bm{C}}_{m,k}^p= \bm{\hat{H}}_{IC,k} \bm{C}_{k+1}^g\bm{\hat{H}}_{IC,k} ^T+\bm{R}_{IC} 
$ ,& \\ 
& $\bm{L}_{k}=chol(\tilde{\bm{C}}_{m,k}^p)
$ ,& \\ 
& $\bm{M}_{k}=\bm{L}_{k}/ \bm{\Delta \hat{y}_k} 
$ ,& \\ 
& $\bm{\mu}_{k+1}^c=\bm{\mu}_{k+1}^g-\bm{C}_{k+1}^g\bm{\hat{H}}_{IC,k}^T (\bm{L}_{k}^T/ \bm{M}_{k} )
$ .& \\ 
\end{tabularx}
\end{center}

\section{Optimizing for Real - Time Evaluation}
\label{sec:realTime}
Even with the Cholesky decompositions, the simultaneous update of all test-grid points might be too slow for real-time implementations. The sequential update that we used in \cite{RGP_mon:2025} was one option to solve this problem. The sequential EKF, however, requires that the measurement noise is uncorrelated. As shown in Sec. \ref{sec:gradient} this marks a simplification for the general case. 
In this paper, we use a similar strategy considering only the pseudo-measurements, i.e., the gradient inequalities w.r.t. the corresponding input dimensions, for a single grid point  per timestep. This simplifies the matrix inversion (or Cholesky decomposition) of an $(\tilde{N} \cdot n_z)\times((\tilde{N} \cdot n_z))$  matrix to an $ n_z\times n_z$ matrix. To ensure that  an update of the whole grid is still conducted, all grid points are updated row-wise under usage of a circular counter. Therefore, the hysteresis heuristic utilized in  \cite{RGP_mon:2025} is not necessary, which should contribute to a better performance in the general case. 

Naturally, this leads to a delayed consideration of the monotonicity knowledge in comparison to the complete update, which is a necessary trade-off.

\subsection{Algorithm}
Since the dimension of the matrix to be inverted has been reduced to $n_z\times n_z$, the computational advantage of a Cholesky decomposition is  small. For better readability we, hence, only present the inversion-based implementation here. If necessary, for evaluation speed or to increase numerical stability, the online algorithm might naturally also be implemented by means of the Cholesky decomposition. 

At timestep $k$ we evaluate the mean values of the gradient  for each direction  $\bm{\mu}_{m,i,k}=\bm{H}_{IC,i} \bm{\mu}_k^g$ and, like in the complete update,  we calculate the gradient difference for each dimension $\bm{\Delta y_{i,k}} =\bm{\mu}_{m,i,k}-B_i $ as well as the corresponding activation matrix $\bm{S}_{i,k}=diag( s_i \bm{\Delta y_{i,k}}>0) $. We introduce a circular counter $o_k$. In timestep $k$, we then loop through the grid points $\tilde{\bm{X}}(o_k,:)$ starting with grid point number $o_{k-1}$ until the monotonicity constraint in at least one direction $i$ is violated. Then, a pseudo-measurement update for this grid point is conducted for all dimensions, according to Sec. \ref{sec:Ineq_Constr}. Naturally, only the active constraints are relevant here. After an update, the "for" loop is broken and in $k+1$, we start testing with the grid point after the one that was updated at last. The "for" loop is employed to avoid an endless loop, if all grid points fulfill the monotonicity constraints in all directions.

This algorithm guarantees that every grid point is considered at least once in every $\tilde{N}$ timesteps but each grid point is taken into account at most once every timestep. Furthermore, only one $n_z\times n_z$-dimensional pseudo-measurement update is conducted, which limits the computational load.

The complete algorithm is summarized in the following subsection.

\subsection{Summary: Complete Monotonicity Update}
Define: 
\begin{itemize}
\item Basis vectors $\bm{X}$ with the expected input bounds $\underline{\zeta}_{i}$ and $\overline{\zeta}_{i}$, the grid points (grid resolution  $N_i$  per input dimension $i$  as described in \eqref{eq:normalization}) and the respective normalization gain $\beta_i$.
\item RGP hyperparameters length-scale $L$, vertical hyperparameter $\sigma_K$, and the measurement noise $\sigma_y$.
\item Monotonicity test vectors $\bm{\tilde{X}}$ with grid resolution  $\tilde{N}_i$  per input dimension $i$  as described in Sec. \ref{sec:gradient} .
\item Pseudo measurement noise $\tilde{\bm{R}}_{IC}$.
\item Desired boundaries $B_i$ and sign - indicator variables for the respective constraints, e.g. $s_i=-1$ for $\frac{\partial z}{\partial \zeta_{i,k}}>B_i$ and $s_i=1$ for $\frac{\partial z}{\partial \zeta_{i,k}}<B_i$.
\end{itemize}

Initialize:
\begin{center}
\begin{tabularx}{1.028\columnwidth}{l X r} 
  \multirow{8}{1em}{\rotatebox{90}{{Offline}}} & $\bm{K}=k(\bm{X},\bm{X}) $ ,  &$\inlineeqno$\\ 
& $\bm{\mu}_{0}^g=\left[\bm{0}\right]^T$ ,&\\ 
& $\bm{C}_{0}^g=\bm{K} $ .&\\ 
& $\bm{H}_{m}$  as given by  \eqref{eq:gradient} and \eqref{eq:meas_matr} .&\\ 
& $\bm{R}_{m,ges}$  as given by \eqref{eq:meas_cov}, \eqref{eq:CrossCov} and \eqref{eq:Cgrad_ges} .&\\ 
& $\bm{R}_{IC}=\bm{R}_{m,ges} +\tilde{\bm{R}}_{IC}$  .&\\ 
& $\tilde{N}=\prod_{i=1}^{n_z}  \tilde{N}_i$ .&\\ 
& $o_0=1$ .&\\ 
\end{tabularx}
\end{center}
Now, evaluate the following recursive algorithm for all steps $k=0,1,2,..$: \newline
1. Complete the RGP Inference  
\begin{center}
\begin{tabularx}{1.028\columnwidth}{l X r} 
  \multirow{4}{1em}{\rotatebox{90}{{Inference}}} & $\bm{\chi}_k=\bm{f}_{norm}(\bm{\zeta}_k) $  ,&$\inlineeqno$\\ 
& $\bm{j}_k^T=k(\bm{\chi}_k^T,\bm{X}) / \bm{K}$ ,& \\ 
& $ \mu_{k}^p=\bm{j}_k^T  \bm{\mu}^c_{k}$ ,&\\ 
& $c_{k}^p=\sigma_K^2+\bm{j}_k^T (\bm{C}_{k}^g-\bm{K}) \bm{j}_k $ .& \\  
\end{tabularx}
\end{center}
2. Calculate the RGP Update
\begin{center}
\begin{tabularx}{1.028\columnwidth}{l X r} 
  \multirow{3}{1em}{\rotatebox{90}{{Update}}} & $\bm{g}_{k}=\bm{C}_{k}^g \bm{j}_k \cdot(c_{k}^p+ \sigma_{y}^2 )^{-1}$ , &$\inlineeqno$ \\ 
& $\bm{\mu}_{k+1}^g= \bm{\mu}_{k}^c+\bm{g}_{k}  (y_{k}- \mu_{k}^p) $,& \\ 
& $\bm{C}_{k+1}^g=\bm{C}_{k}^g-\bm{g}_{k} \bm{j}_k^T\bm{C}_{k}^g \,.$ & \\ 
\end{tabularx}
\end{center}
3. Perform the Monotonicity Update
\begin{center}
\begin{tabularx}{1.028\columnwidth}{l X r} 
  \multirow{19}{1em}{\rotatebox{90}{{Monotonicity Update}}}&  $\bm{\mu}_{m,i,k}=\bm{H}_{IC,i} \bm{\mu}_{k+1}^g$ ,&$\inlineeqno $ \\
& $\bm{\Delta y_{i,k}} =\bm{\mu}_{m,i,k}-B_i $ ,& \\ 
& $\bm{S}_{i,k}=diag( s_i \bm{\Delta y_{i,k}}>0)$ ,& \\ 
& $\hat{\bm{H}}_{IC,i,k}=\bm{S}_{i,k} \bm{H}_{IC,i}$ ,& \\ 
& $\bm{\Delta \hat{y}_{i,k}} = \bm{S}_{i,k} \bm{\Delta y_{i,k}}$ ,& \\
& $o_k=o_{k-1}$ ,& \\ 
& $\mathrm{For}~ l=1, \dots, \tilde{N}$: & \\
& ~~ $\bm{\hat{H}}_{IC,k,o}=\left[\bm{\hat{H}}_{IC,1,k}(o_k,:)^T,\bm{\hat{H}}_{IC,2,k}(o_k,:)^T,\dots\right]^T$ ,& \\ 
& ~~ $\bm{R}_{m,ges,o}=diag([\bm{R}_{m,ges}(o_k,o_k),\bm{R}_{m,ges}(o_k+1\tilde{N},o_k+1\tilde{N}),\dots])$ ,& \\ 
& ~~ $\bm{R}_{IC,o}=diag([\bm{R}_{IC}(o_k,o_k),\bm{R}_{IC}(o_k+1\tilde{N},o_k+1\tilde{N}),\dots])$ ,& \\ 
& ~~ $\mathrm{if}~ \bm{S}_{1,k}(o_k,o_k)>0~ or ~\bm{S}_{2,k}(o_k,o_k)>0, \dots$: & \\ 
& ~~~~~~  $\bm{\Delta \hat{y}_{k,o}}=\left[\bm{\Delta \hat{y}_{1,k}}(o),\bm{\Delta \hat{y}_{2,k}}(o),\dots \right]^T$ ,& \\
& ~~~~~~  $\bm{\mu}_ {k+1}^c=\bm{\mu}_{k+1}^g-\bm{C}_{k+1}^g\bm{\hat{H}}_{IC,k,o}^T \left(\bm{\hat{H}}_{IC,k,o}\bm{C}_{k+1}^g\bm{\hat{H}}_{IC,k,o} ^T+\bm{R}_{m,ges,o}+ \bm{R}_{IC,o} \right)^{-1} \bm{\Delta \hat{y}_{k,o}}$, & \\ 
& ~~~~~~  $o_k=o_k+1$ ,& \\ 
& ~~~~~~  $\mathrm{if}~ o_k>\tilde{N}: o_k=1$ ,& \\ 
& ~~~~~~  break  loop ,& \\ 
& ~~ $\mathrm{else}~$: & \\ 
& ~~~~~~  $o_k=o_k+1$ ,& \\ 
& ~~~~~~  $\mathrm{if}~ o_k>\tilde{N}: o_k=1$ .& \\ 
\end{tabularx}
\end{center}

\section{Numerical Validation}
In this section, we will numerically verify the presented methods enforcing monotonicity constraints in RGP, to which we will refer to as RGPm. Therefore, we will show how they compare against the standard RGP method. Furthermore, we compare them with a selection of methods from \cite{RGP_mon:2025} which aimed at the same goal.

\subsection{1D Simulation Example}
\begin{figure}
	 \begin{center}
		 \includegraphics[width=1\linewidth]{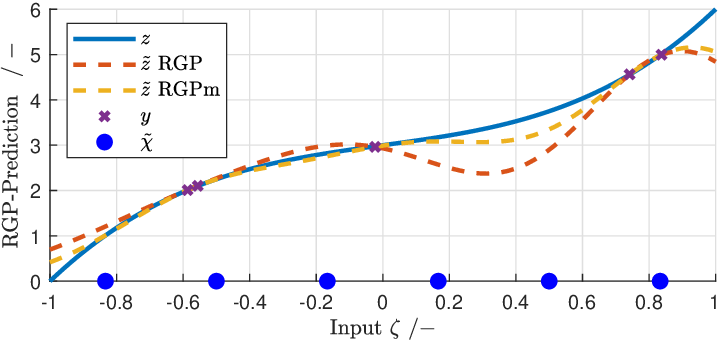}
		  \caption{RGP and RGPm outputs in comparison with the hidden function $z$ after 5 timesteps and the utilized test grid.} 
		  \label{pic:BSP_mean}
	 \end{center}
\end{figure}

To provide a qualitative indication of the functionality of our algorithm, we first compare an exemplary simulation for a  1D hidden function after 5 measurements for the basic RGP and RGPm (here with an exact, simultaneous update, see Sec. \ref{sec:EKF}) in Fig. \ref{pic:BSP_mean}. It becomes obvious that due to the consideration of the covariance, the monotonicity update only alters the solution in regions with high uncertainty, whereas RGP and RGPm produce nearly identical results in the vicinity of the measurements. It also becomes clear that the monotonicity updates are only executed on the test grid. If the test grid is not fine enough, there might be regions where the constraints are not satisfied.

\subsection{2D Statistical Validation}
In the following, we present the results for  a 2-dimensional hidden function $z=10 \zeta_{1,k}+1 \zeta_{1,k}^3+ 10 \zeta_{2,k}$. The inputs $\bm{\zeta}_k=\left[\zeta_{1,k},\zeta_{2,k}\right]^T$ are picked from two random uniform distributions over the complete input range $\zeta_{1,k} \sim \mathcal{U}(-2,4)$ and  $\zeta_{2,k} \sim \mathcal{U}(-1,4)$. Moreover, zero-mean Gaussian white noise with a variance of  $\sigma_y^2=1e-1$ is added to the measured output $y_k$.

The RGP hyperparameters are chosen as follows: $L=1.5$, $\sigma_K=1e1$, and $N_1=N_2=10$. Obviously, the hidden function is strictly monotonically increasing in both dimensions, so $\frac{\partial z}{\partial \zeta_{1,k}}>0$ and $\frac{\partial z}{\partial \zeta_{2,k}}>0$ ,  $B_1=B_2=0$, and thus $s_1=s_2=-1$ hold. The pseudo-measurement noise is chosen as $\bm{\tilde{R}}_{IC}=1e-2\cdot \bm{I}$.

For a statistical validation of the algorithm and an assessment of the impact of the adaptations, we investigate five different variants of the algorithm for two different test grid sizes:
\begin{itemize}
\item[S0] Pure RGP
\item[S1] Benchmark RGPm from \cite{RGP_mon:2025} with unlimited  $\tilde{n}_{IC}=n_{IC}$  with  $\delta_b=0$ and $\delta_u=0$ 
\item[S2] Benchmark real-time optimized RGPm from \cite{RGP_mon:2025} with $\tilde{n}_{IC}=2$ and  $\delta_b=1e-1$ and $\delta_u=1e-1$ 
\item[S3] RGPm with exact update, see Sec. \ref{sec:Ineq_Constr} 
\item[S4] Real-time optimized RGPm, see Sec. \ref{sec:realTime} .
\end{itemize}

For each algorithmic variant, 500 simulation runs are conducted with the described uniform random input and the noisy output. After $k=1,2,5, \dots ,1000$ steps, the root mean squared error (RMSE) between the learned function of each variation, compared to the actual hidden function, is calculated for an equidistant evaluation test grid, that covers the complete input range of the function (e.g. $\underline{\zeta}_i  \dots\overline{\zeta}_i$). This RMSE is again averaged over all 500 simulations and depicted in Fig. \ref{pic:Sim_Stat_RSME_low} and Fig. \ref{pic:Sim_Stat_RSME_high}  for the respective test grid resolutions, i.e., $\tilde{N}_1=\tilde{N}_2=5$ and $\tilde{N}_1=\tilde{N}_2=10$. An overview over the hyperparameters is given in Table \ref{table:Hyperparamters_sim}.

  \begin{table}
\centering
  \caption{Parameters and Hyperparameters for the Simulation}
\begin{tabular}{ c c c c } 
  \multicolumn{4}{c}{RGP Hyperparameters}\\
\hline
 $n_z$  & $2$ & $N_1$  & $10$ \\
 $N_2$  & $10$ & $\underline{\zeta}_1$  & $-2$  \\
$\overline{\zeta}_1$  & $4$ & $\underline{\zeta}_2$  & $-1$  \\
$\overline{\zeta}_2$  & $4$  &  $L$ & $1.5$ \\
$\sigma_K$  & $1e1$  &  $\sigma_y$ & $1e-1$ \\
 \hline
  \multicolumn{4}{c}{Monotonicity Hyperparameters}\\
\hline
 $s_1$  & $-1$ & $s_2$  & $-1$ \\
  ${B}_1$  & $0$ & ${B}_2$  & $0$  \\
$\tilde{\bm{R}}_{IC}$  & $(1e-1)^2  \bm{I}$  &   & \\
\hline
  \multicolumn{4}{c}{Coarse Test Grid}\\
\hline
$\tilde{N}_1$& $5$  & $\tilde{N}_2$& $5$ \\
  \multicolumn{4}{c}{Fine Test Grid}\\
\hline
$\tilde{N}_1$& $10$  & $\tilde{N}_2$& $10$ \\
 \hline
  \end{tabular}%
\label{table:Hyperparamters_sim}
\end{table}

\subsection{2D Statistical Validation: Results}
\begin{figure}
	 \begin{center}
		 \includegraphics[width=1\linewidth]{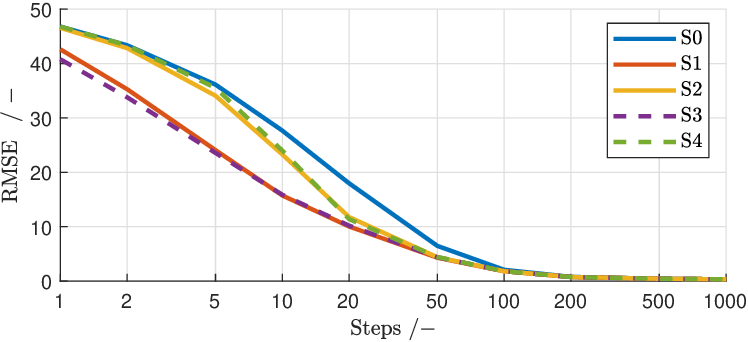}
		  \caption{Average RMSE for 500 simulation runs of the different RGP and RGPm variants with a fine test grid.} 
		  \label{pic:Sim_Stat_RSME_low}
	 \end{center}
\end{figure}

In Fig. \ref{pic:Sim_Stat_RSME_low}, the RMSEs for the coarse test grid for the alternative variants are depicted. It is obvious that all RGPm variants improve the RGP-baseline performance further. This improvement has the highest impact at the beginning and declines with steps and more available data. This is to be expected, since the additional knowledge introduced by assuming monotonicity is also represented in the data. The algorithm leveraging the exact gradient covariance (S3) performs only slightly better than S1, and is similar in terms of the computational effort, as can be seen in Table \ref{table:Evaltime_Sim}. A similar comparison can be drawn between the offline variants from this paper (S4) and S2. Here, S2 performs slightly better. It might be the case that the hysteresis-like heuristic has a positive effect here, or that the two update steps in S2 are beneficial in comparison with the single update step in S4. The computation time depicted in Table \ref{table:Evaltime_Sim} is also almost identical. 

\begin{figure}
	 \begin{center}
		 \includegraphics[width=1\linewidth]{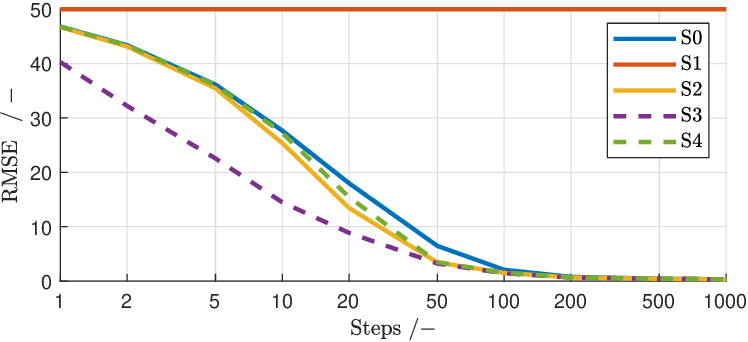}
		  \caption{Average RMSE for 500 simulation runs of the different RGP and RGPm variants with a coarse test grid, limited to $50$.} 
		  \label{pic:Sim_Stat_RSME_high}
	 \end{center}
\end{figure}

The results for the fine test grid in Fig. \ref{pic:Sim_Stat_RSME_high} show quite different results. Here, S1 leads to a diverging RGP prediction. The depiction in the plot is limited to 50, but true values approach infinity. The update based on exact covariance S3, on the other hand, provides very good results, slightly improving on the coarser test grid as  expected. The evaluation speed of S3, depicted in Table \ref{table:Evaltime_Sim}, is also significantly faster, which is most likely caused by the efficient implementation by means of the Cholesky decomposition.  Both online methods S2 and S4 a slightly worse performance than for the coarser grid. This is most likely due to the fact that it takes longer to iterate through all the grid points so that regions of the hidden function  longer remain unconsidered w.r.t. to the monotonicity constraints. Whether this negative effect or the positive effect of a finer grid will be more influential depends most likely on the RGP measurement noise $\sigma_y$ and the spread of available data. With high noise, or if data is not available for the whole input range of the RGP, the benefit of a finer grid will most likely dominate, even for the online methods. The variant S4 is slightly slower here than S2 but still a lot faster than the full grid evaluation. 

At a desktop PC with an Intel i9-14900KF CPU and 128 GB RAM, the RGP-baseline took about $90~\mu s$. The maximum evaluation times for a real-time hardware, which might be more relevant, is given in Sec. \ref{sec:ex_val}.

  \begin{table}
\centering
  \caption{Mean evaluation times, normalized to the RGP-baseline}
\begin{tabular}{ c | c | c  } 
RGPm variant & fine test grid & coarse test grid \\
\hline
S0 & 1 & 1 \\
S1 & 2.2 & 19.2 \\
S2  & 1.2 &     1.5 \\
 S3  & 2.4 & 6.4\\
S4   & 1.2 &     1.8 \\
 \hline
  \end{tabular}%
\label{table:Evaltime_Sim}
\end{table} 

\subsection{Discussion}
In general, the divergent behavior of the RGPm algorithm from \cite{RGP_mon:2025} was only observed with a fine test grid. Since S1 does not provide any advantage in terms of computational speed, the full update with exact covariances (S3) is the recommended method if the computational power allows for it. With a low limit of updates per timestep, the online-variant from \cite{RGP_mon:2025} (S2) did not show the divergent behavior in any of the conducted tests. It possesses the theoretical disadvantages that were covered in Sec. \ref{sec:realTime} but behaves slightly superior in terms of performance and computational speed then the online variant presented in this paper. In general, we recommend S3 if enough computational power is available, otherwise S4. S2, on the other hand, should be employed only after a thorough validation for the respective application scenario.

\section{Experimental Validation with a Pneumatic Control Application}
\label{sec:ex_val}
\begin{figure}
	 \begin{center}
		 \includegraphics[width=0.6\linewidth]{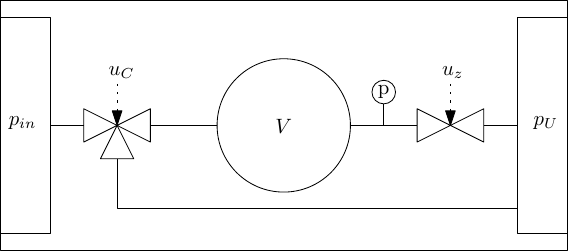}
		  \caption{Scheme of the considered pneumatic test rig at the Chair of Mechatronics, University of Rostock.} 
		  \label{pic:PneumPruefstand}
	 \end{center}
\end{figure}

In this chapter, we present an experimental validation of the RGPm algorithm. In the experimental validation of the original RGPm algorithm in \cite{RGP_mon:2025}, we focused on a Vapor Compression Cycle (VCC) as a control application, which originally motivated the algorithmic developments. While surely marking a relevant application, the complexity of that system is not easy to grasp and might inhibit the understanding and, moreover, its large time constants would increase the necessary experimental time for a statistically viable analysis. Furthermore, it is very difficult to achieve repeatable environmental conditions for the VCC, which complicates the comparability of the experiments. Consequently, we decided to consider the much simpler pneumatic control application depicted in  Fig.~\ref{pic:PneumPruefstand} for an experimental validation of the algorithms proposed in this paper. The system consists of two pneumatic valves, a pneumatic tank with a storage volume $V=4e-4~\mathrm{m^3} $ and a pressure sensor. The first pneumatic valve represents a 5/3 way valve, which is able to fill or deplete the tank based on the supply pressure $p_{in}$  or the ambient pressure $p_U$, respectively. The second valve is only employed as a throttle with variable diameter. The first valve possesses a known characteristic, and its voltage is the control input $u_C$, whereas the second valve characteristic is unknown but the corresponding control voltage $\zeta_{1,k}=u_z(k)$ represents a measurable disturbance. The control system is implemented on a Bachman PLC (CPU:MH230) with a sample time of $T_s=1~ms$.

\begin{figure}
	 \begin{center}
		 \includegraphics[width=0.8\linewidth]{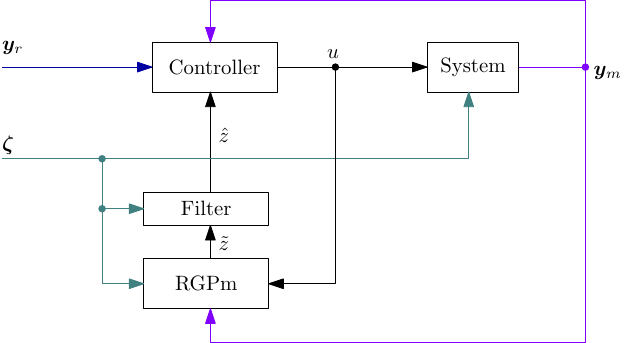}
		  \caption{Control structure for the experimental validation.} 
		  \label{pic:ControlStructure}
	 \end{center}
\end{figure}

The control structure depicted in Fig. \ref{pic:ControlStructure} consists of a model-based controller, the RGPm algorithm and a filter. The RGPm algorithm uses the steady state values of the model-based control for learning. The RGP-model prediction is filtered and then used within the control. The special filter, which shall not be derived here in detail, is at its core a rate limiter that is switched on and off depending on the current prediction uncertainty. It is necessary to preserve stability of the overall control structure for a simultaneous learning and application of the RGP or RGPm model.

\subsection{Model-Based Controller}
The pneumatic storage volume can be modelled by a mass-balance
\begin{equation}
\frac{dm}{dt}=\dot{m}_{in}-\dot{m}_{out} \label{eq:Mbalance} \,.
\end{equation}
The air mass inside the tank is related to the tank pressure by means of the ideal gas law
\begin{equation}
\frac{p V}{R T} =m \,.
\end{equation}
Assuming an isothermal thermodynamic process, the time derivative becomes
\begin{equation}
\frac{\dot{p} V}{R T} =\dot{m} \,
\end{equation}
and can be substituted in \eqref{eq:Mbalance}
\begin{equation}
\frac{dp}{dt}=\frac{R T}{V} \left(\dot{m}_{in}-\dot{m}_{out} \right) \label{eq:ODEpneum} \,.
\end{equation}
The input mass flow depends in a nonlinear manner on the pressure drop over the first valve and its actuation voltage $\dot{m}_{in}=g(p_{in},p_{U},p,u_C)$. This relation and also its inverse $u_C=g_{I}(p_{in},p_{U},p,\dot{m}_{in})$ are known from a system identification. The output mass flow $\dot{m}_{out}=z(p,p_U,u_z)$ through the second valve, which depends on the corresponding pressure drop and the measurable actuation voltage, is unknown and represents the disturbance.  In our test scenario, this valve actuation is chosen as a random signal. In a real application, it may stem from a secondary controller. 

The overall nonlinear system dynamics can be represented in the following first-order ODE
\begin{equation}
\frac{dp}{dt}=\frac{R T}{V} \left(g(p_{in},p_{U},p,u_C)-z(p,p_U,u_z) \right) \,.
\end{equation}
A well-established method for the control of pneumatic systems, as proposed in \cite{Wache:2019}, is an inversion of the input nonlinearities by evaluating $u_C=g_{I}(p_{in},p_{U},p,\dot{m}_{in,d})$ and defining $\dot{m}_{in,d}$ as a new input. This technique is applied here as well, which results in a disturbed linear system with state $x=p$
\begin{equation}
\frac{dx}{dt}=\underbrace{\frac{R T}{V}}_b \dot{m}_{in,d}+\underbrace{-\frac{R T}{V}}_e z(p,p_U,u_z)  \,.
\end{equation}
This system is now controlled with an IOL tracking controller that comprises an integral feedback term to achieve steady-state accuracy. The overall control law results in the stabilizing feedback
\begin{equation}
\upsilon=\dot{x}_r +k (x_r-x)+k_I\int (x_r-x) dt
\end{equation}
and the inverse dynamics
\begin{equation}
\dot{m}_{in,d}=\frac{1}{b} \upsilon- \frac{e}{b} \hat{z},
\end{equation}
evaluated with $u_C=g_I(p,p_U,p_{in},\dot{m}_{in,d})$ and the filtered output $\hat{z}$ of the learned disturbance model $\tilde{z}$.

\subsection{RGPm Implementation}
To reduce the input dimension, we combine the first two disturbance inputs in the pressure factor $p_{fac}=\frac{p_U}{p}$. Consequently, the actual disturbance model to be learned with the RGPm algorithm is given by $\tilde{z} (p_{fac},u_z)$.  For numerical reasons, the mass flow is defined in the unit $g/s$. 

From the model equations \eqref{eq:ODEpneum}, we know that in steady state $\dot{m}_{out}=\dot{m}_{in}$ must hold. Thus, we use $y=\dot{m}_{in}$ as  the measurement output for the RGP. We detect the vicinity to the steady state by checking whether  $|x_r-x|<Lim_1$ is below a defined margin. Otherwise, we disable learning. Generally, $|\dot{x}_r|<Lim_2$ must checked as well to detect a steady state. For the stepwise trajectory, this additional test has, however, no influence on the results. 

 As known from our own experiments and from the literature, the inequalities $\frac{\partial {z}}{\partial p_{fac}}<0$  and $\frac{\partial {z}}{\partial u_z}<0$ hold. Consequently, they are included as monotonicity assumptions in the RGPm algorithm. The evaluation of the RGP model is conducted with the reference value $p_{fac,r}$ instead of the current one, i.e., $\hat{z} (p_{fac,r},u_z)$. This reduces the influence of noisy measurements.

The RGP and RGPm algorithms are implemented with basis vector dimensions of $N_1=5$ and $N_2=5$ for the respective RGP inputs, which results in $N_X=25$. The monotonicity test grid is equally defined with $\tilde{N}_1=\tilde{N}_2=5$. Moreover, the online update from Sec. \ref{sec:realTime} is utilized. The pseudo-measurement noise is characterized by $\tilde{\bm{R}}_{IC}=1e-1^2 \cdot \bm{I}$, and the length scale is set to $L=1.5$. The signing variables for the previously mentioned monotonicity assumptions are given by $s_1=-1$ and $s_2=-1$, and the safety margins are disabled, i.e., $B_1=B_2=0$.

  \begin{table}
\centering
  \caption{Parameters and Hyperparameters for the Experiments}
\begin{tabular}{ c c c c } 
\hline
\multicolumn{4}{c}{System and Control Parameters}\\
\hline
 $T_s$  & $1e-3~s$ & $R$  & $0.2871 \frac{J}{g K}$ \\
 $T_0$  & $293.15 K$ & $V$  & $4e-4~m^3$  \\
 $k_C$  & $4$  &  $k_I$  & $4$ \\
 $\sigma_{fac}$  & $5$  &  $ $ &  \\
 \hline
 \multicolumn{4}{c}{RGP Hyperparameters}\\
\hline
 $n_z$  & $2$ & $N_1$  & $5$ \\
 $N_2$  & $5$ & $\underline{\zeta}_1$  & $0~\frac{g}{s}$  \\
$\overline{\zeta}_1$  & $5~\frac{g}{s}$ & $\underline{\zeta}_2$  & $0$  \\
$\overline{\zeta}_2$  & $1$  &  $L$ & $2.5$ \\
$\sigma_K$  & $1e0$  &  $\sigma_y$ & $1e1$ \\
 \hline
  \multicolumn{4}{c}{Monotonicity Hyperparameters}\\
\hline
$\tilde{N}_1$ & $5$  & $\tilde{N}_2$ & $5$ \\
 $s_1$  & $1$ & $s_2$  & $1$ \\
  ${B}_1$  & $0$ & ${B}_2$  & $0$  \\
$\tilde{\bm{R}}_{IC}$  & $(1e-1)^2  \bm{I}$  &   & \\
 \hline
  \end{tabular}%
\label{table:Hyperparamters_ex}
\end{table}

\subsection{Test Setup for the Experimental Validation}
\begin{figure}
	 \begin{center}
		 \includegraphics[width=1\linewidth]{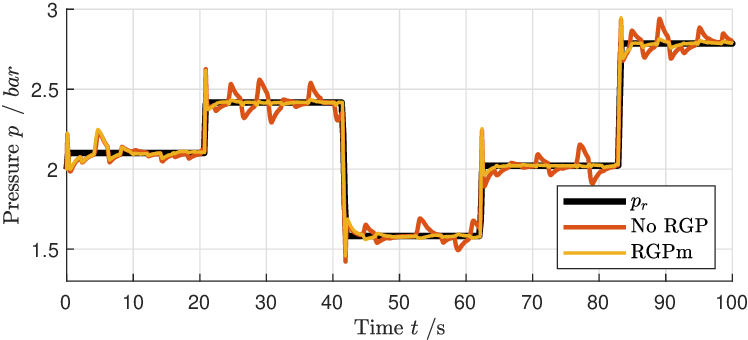}
		  \caption{Experimental results for one exemplary validation test run.} 
		  \label{pic:y_dyn}
	 \end{center}
\end{figure}

To reduce the impact of stochastic effects, the validation is structured as follows. A low-pass filtered sequence of steps serves as reference trajectory for the desired pressure as shown in Fig.~\ref{pic:y_dyn} , which is employed for 5 runs with $100~s$ each. Within this time span, equally random and filtered stepwise trajectories with a higher sample time  are defined for $u_z$. All 5 runs are evaluated for all variants and algorithms with a settling time of $20~s$ between the runs. Moreover, the RGPs are re-initialized after all single runs. 

\subsection{Experimental Results}
\begin{figure}
	 \begin{center}
		 \includegraphics[width=1\linewidth]{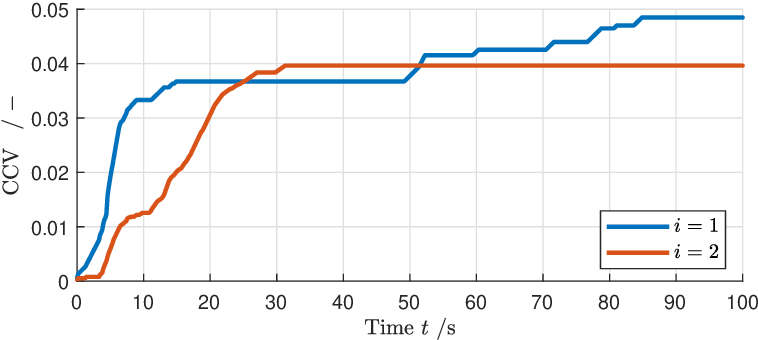}
		  \caption{Cumulated constraint violations (CCV) over 5 runs for the 2 dimensions normalized to the RGP baseline.} 
		  \label{pic:CumulConstrViol}
	 \end{center}
\end{figure}

In Fig. \ref{pic:CumulConstrViol}, we depict the cumulated constraint violation for the 2 dimensions normalized to the RGP baseline. As we can see, the RGPm algorithm reduces the  constraint violations for both dimensions by more then $95~\%$. Furthermore, most of the violations occur at the beginning. Ultimately, this proves the effectiveness of the algorithm in an experimental setup. 

\begin{figure}
	 \begin{center}
		 \includegraphics[width=1\linewidth]{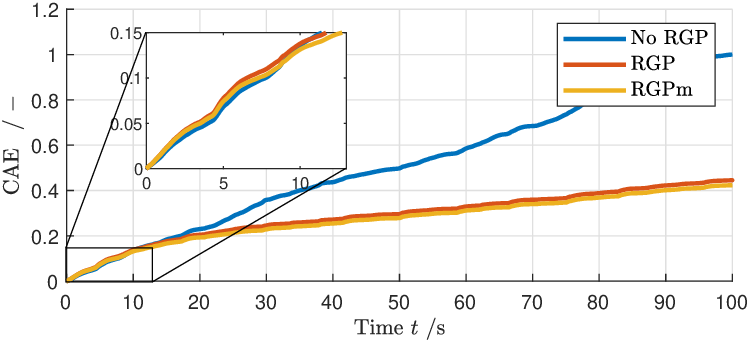}
		  \caption{Cumulated Average Error (CAE) over 5 runs normalized to the RGP baseline for different control variants.} 
		  \label{pic:CumError}
	 \end{center}
\end{figure}

Fig.~\ref{pic:y_dyn} already indicates the general performance increase by the use of the RGPm algorithm in the control structure. As can be seen in the cumulative absolute error (CAE) normalized to the baseline in Fig. \ref{pic:CumError}, there is a drastic improvement  of about $55~\%$ compared to the baseline, after a short period where the RGP and RGPm algorithm slightly worsen the performance. This short period at the beginning can be explained by the fact that the RGP model, which is applied in the control structure, is still highly uncertain in this early learning period. While the overall improvement of the RGPm algorithm compared to the RGP algorithm is only about $2~\%$, it is especially effective in this starting region of the first $20~s$ as was already noticed in the numerical validation in Sec.~\ref{sec:simul_eval}.

\begin{figure*}[h!]
	 \begin{center}
\includegraphics[width=0.49\linewidth]{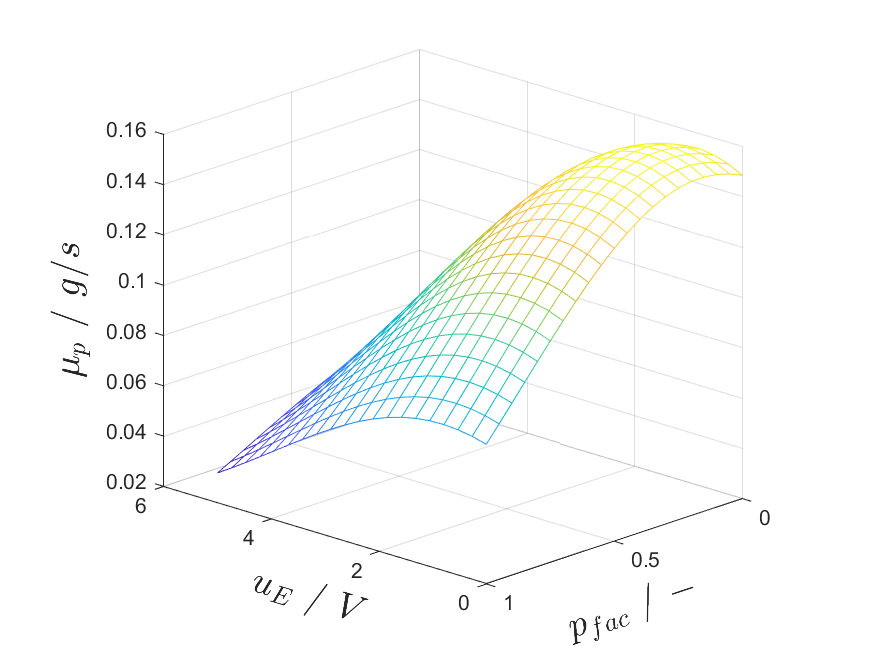} 
\includegraphics[width=0.49\linewidth]{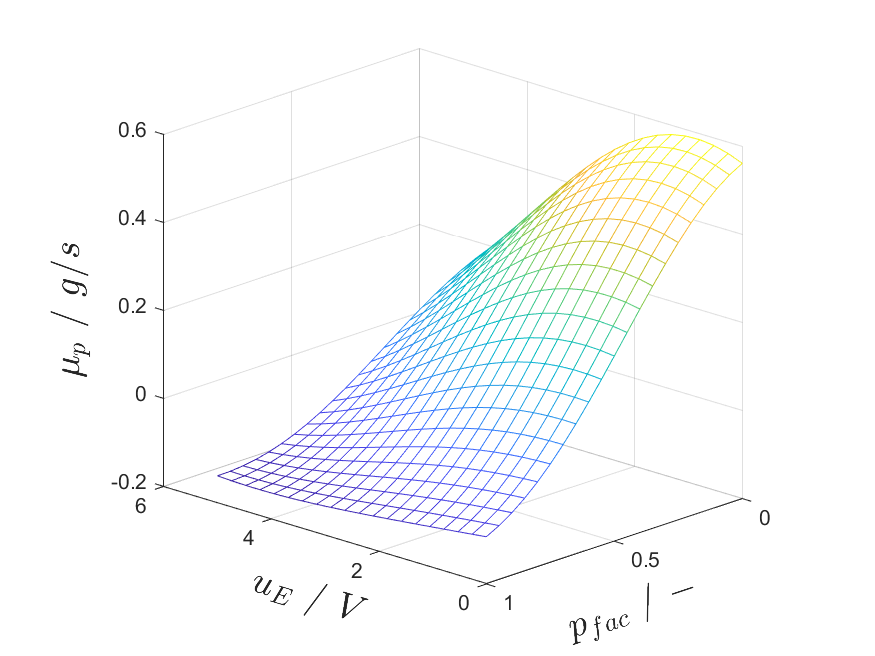}
\includegraphics[width=0.49\linewidth]{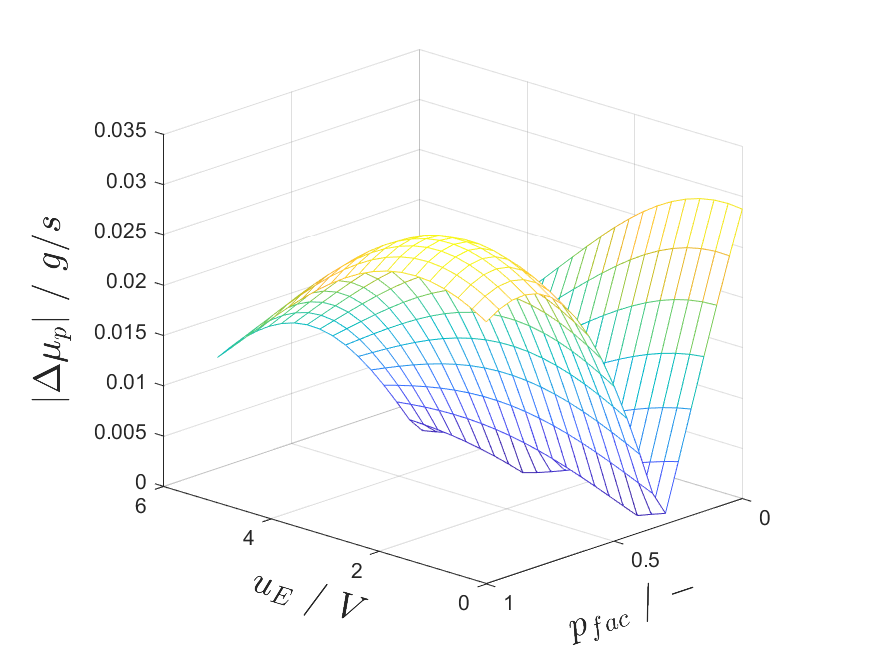} 
\includegraphics[width=0.49\linewidth]{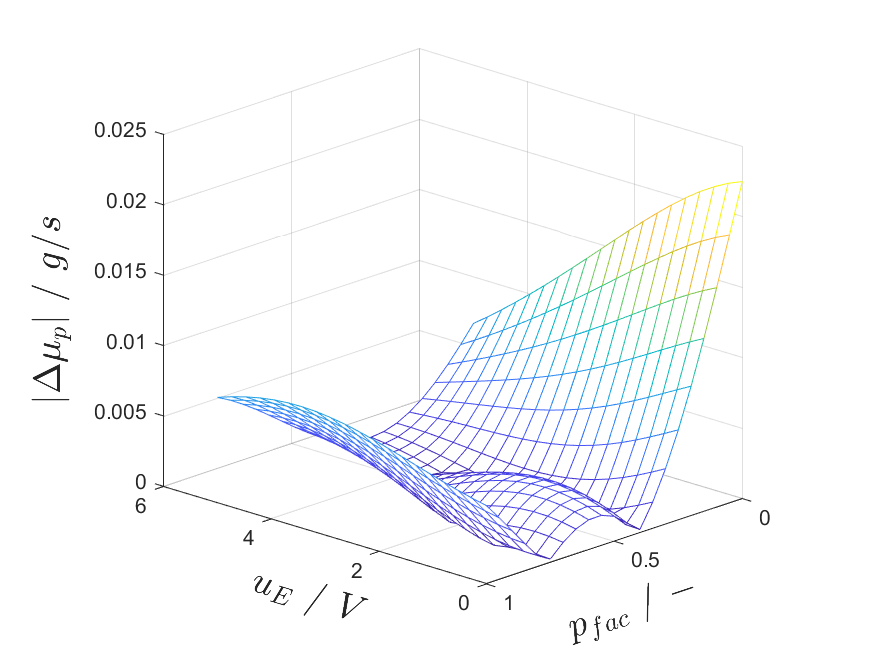}
\includegraphics[width=0.49\linewidth]{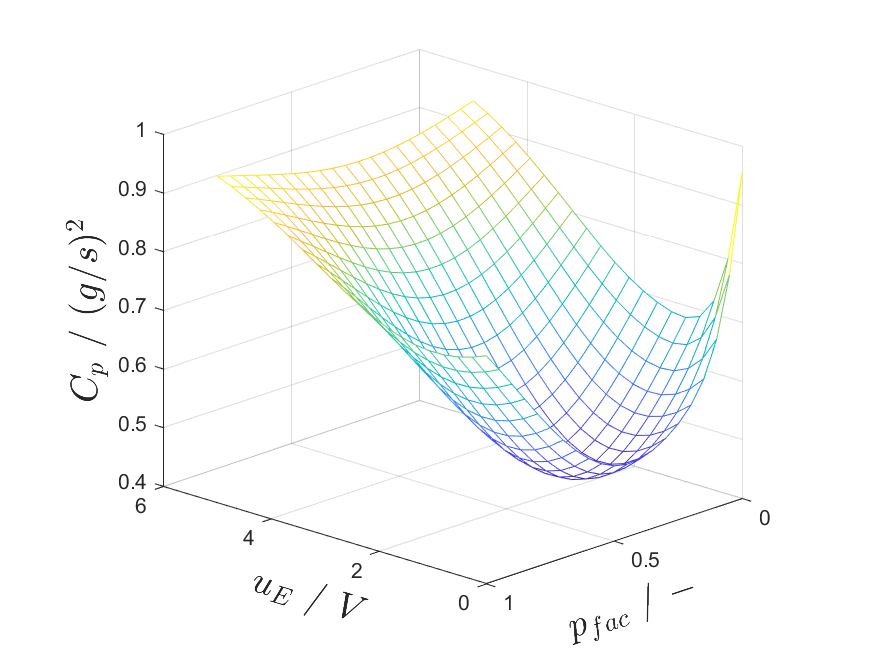} 
\includegraphics[width=0.49\linewidth]{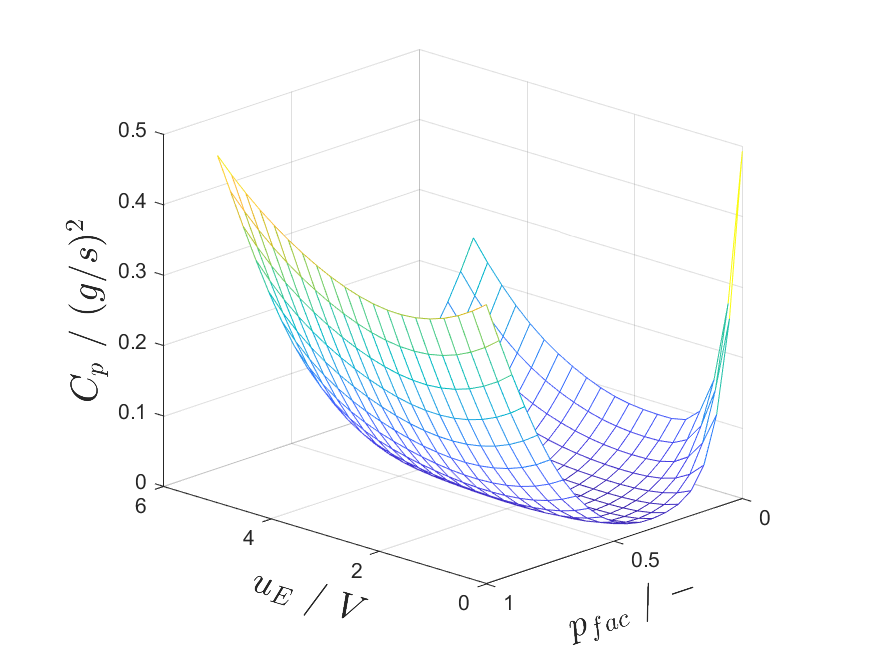}
		  \caption{RGP model predictions from top to bottom for the same disturbance and reference trajectories: RGPm , RGPm-RGP absolute difference and covariances for RGPm.  From left to right: after $1~s$ and $100~s$ runtime. } 
		  \label{pic:DreiDplot}
	 \end{center}
\end{figure*}

In the Fig. \ref{pic:DreiDplot}, we depict the RGP predictions of the RGPm algorithm as well as the deviation to the RGP algorithm and  the covariance after $1~s$ and $100~s$, respectively. For RGP and RGPm, the same $p_d$- and $u_z$- trajectories were used with a length of $100~s$. Hence, they were both trained with roughly the same data. This is obvious when comparing the covariances, which had a maximum deviation of $0.0026$. Hence, only the RGPm covariance is depicted. 

The resulting predictions clearly indicate that the RGPm model behaves in accordance with the monotonicity constraints. Furthermore, RGP and RGPm deviate mostly in regions where the covariance is high because only a little amount of data is available. This is consistent with the desired behavior of the RGPm algorithm to change the RGP model only in regions of high uncertainty. Finally, the deviations between RGP and RGPm model generally decrease at the later point in time, where the available data provides a consistent model, also for the RGP. These results are similar to the numerical validation of Sec. \ref{sec:simul_eval} and the CAE from Fig. \ref{pic:CumError}, which make clear, that the RGPm algorithm is especially beneficial when only a little amount of data is available.

In Table. \ref{table:Evaltime_Ex}, we provide an overview over the maximum evaluation time of the different control variants over the experiments. For the given hyperparameters, the RGP algorithm increases the baseline by over $300~\mu s$, whereas while the monotonicity update only leads to a further increase of $60~\mu s$, so that it does not cost much in terms of computation. Please note that these numbers shall only provide a rough runtime estimate since there is still a lot of computational overhead in the code for data-logging and evaluation purposes.

  \begin{table}
\centering
  \caption{Maximum Evaluation Time}
\begin{tabular}{ c c  } 
Variant & $\mu s$\\
\hline
No RGP & 84  \\
RGP  & 411 \\
 RGPm  & 472 \\
 \hline
  \end{tabular}%
\label{table:Evaltime_Ex}
\end{table}

\section{Conclusions and Outlook}
This paper presents an extension of the recursive Gaussian Process regression (RGP) algorithm to enforce (soft) monotonicity constraints during an online training. Therefore, we introduce a runtime-optimized algorithm which utilizes the exact RGP-gradient prediction for a given test grid. Furthermore, an even more accelerated version of this algorithm is presented, which updates the test grid points over consecutive timesteps. The algorithms are validated and compared to previously published work for a numerical 2D example. The real-time optimized version of the algorithm is then successfully experimentally validated for a pneumatic system in combination with a model-based controller. 

A combination of the extension presented in this paper with the Kalman Filter integration of the RGP (GPSOL or RGP-dKF, see \cite{RGP_dKF:2025})  is straightforward and allows for an RGP training with monotonicity constraints if the output of the hidden function is not directly measurable. For systems where a hard constraint satisfaction is crucial, further investigations are still necessary. A viable method could include an additional safety step validating the constraints on a finer grid.

%

%
%
%
 \bibliographystyle{splncs04}
\bibliography{myrefs}

\end{document}